# Design of a 10 picosecond Time of Flight Detector using Avalanche Photodiodes

Sebastian White[a,*], Mickey Chiu[a], Milind Diwan[a], Grigor Atoyan[b] and Vladimir Issakov[b]

[a] Brookhaven National Laboratory, Upton, NY 11973, USA

[b] Yale University, PO Box 208121, 554JWG, New Haven, Ct. 06520

**Abstract**

We describe a detector for measuring the time of flight of forward protons at the Large Hadron Collider (LHC) up to and beyond the full instantaneous design luminosity of $10^{34}$ cm$^{-2}$ s$^{-1}$. Our design is based on deep diffused, high gain avalanche photodiodes (APDs) which give a signal of ~6000 $e$ when traversed by 7 Teravolt protons. We observe pulse risetimes of 650 psec and a pulse width of 5 nsec and calculate a time resolution of ~10 psec and a maximum count rate of >100 Mhz.

**1. Time of Flight with APDs:**

We propose to use fast timing in conjunction with small silicon tracking detectors over an area of 2 cm$^2$. When the detectors are inserted into the beam pipe using a Roman Pot, space is very limited.

In this paper we show that, for charged particle tracking over regions of a few cm$^2$, particularly at high rates, APDs are a natural solution for high precision time of flight (TOF).

There are many examples of APDs as photodetectors used for timing [1].but not as charged particle detectors.

A new fast Hybrid photodetector [2] uses an APD to measure charged particle timing. In it photoelectrons are accelerated to 8kV and lose energy in an APD target providing a total signal of $10^5$ electrons after an internal multiplication of $10^2$. The time resolution due to the APD is found to be 10 psec.

In ref. [3] they were also used as particle detectors. The time resolution of EG&G Avalanche Diodes (AVDs) was measured with ~3 MeV β s. Using an internal gain of 45, a signal risetime of 2.5 nsec and a preamplifier dominated noise of SNR=42 they found a time resolution of 65-87 psec.

Large amplitude variations were observed due, mostly, to fluctuations in the number of electron-hole pairs produced by ionization energy loss in the active silicon layer. The resulting time walk was reduced by a factor of >10 taking into account the measured pulse height.

In the devices we consider, the risetime is ~650 psec and the amplitude fluctuations are σ~18% so, with a similar level of walk correction, we expect a time resolution of 10psec. The jitter due to noise is smaller than this.

**2. Application:**

Recently there has been considerable interest in the central exclusive production process at the LHC- i.e.

$$pp \rightarrow p + X^0 + p \qquad (1)$$

where $X^o$ could be a new particle such as the Higgs boson. Here "+" denotes a region completely devoid of activity, usually referred to as a rapidity gap. The final state therefore consists only of the decay products of the particle $X^o$ and the two outgoing forward protons.

It is very useful to measure the momenta of the two protons in the forward direction. Using their momenta and conservation of energy, the mass of $X^0$ can be measured to ~2-3% irrespective of its decay mode. Since in (1) the quantum numbers of $X^0$ can be determined, this measurement could be important for establishing that it is the Higgs particle.

In the case of the Higgs boson the production cross section is so small that the experiment must run at the maximum LHC intensity. Full intensity is, in any case, required for the program of the rest of the experiment.

Then, because of the bunched structure of the beams, ~25 collisions occur within an r.m.s. time of 170 psec every 25 nsec.

At the locations 420 m from the ATLAS or CMS experiments 1% of these interactions produce a leading proton entering the forward detector area.

So the expected signal rate is 1%*1 GHz =10 MHz in an area extending +/-5 mm vertically and 20 mm horizontally. The background rate and total dose for all species of particles have been calculated [4] for both this location and 220 m discussed below. In the 420 m case we find a total rate of 20Mhz including both protons and electrons from beam loss. The dose is dominated by the protons.

**3. Requirements:**

*3.1. Timing*

Because of the large number of interactions within a bunch, combinatorial background is a serious problem at full intensity.

This can be reduced with proton timing since the time difference of the two protons determines the position of the interaction along the beam. The time average can also be correlated with that of particles in the central detector.

A detailed analysis [5][6] shows that the required time resolution of the Time of Flight detector[TOF] is 10 psec for acceptable reduction of the combinatorial diffractive background at instantaneous luminosities of $10^{34}$ cm$^{-2}$ s$^{-1}$

In ref. [7] a time resolution of <10 psec was obtained with a Cerenkov radiator and a Hamamatsu R3809U-50 Micro channel plate-PMT (MCP-PMT) as the photodetector. We are concerned about the rate limitations of existing MCP-PMTs since, extrapolating the running conditions in [7] to 20 MHz for $10^7$ sec, we find an integrated anode current of

#photoelectrons*Gain*Rate*30%*1 year*$q_e$

= >70*10⁶*20*10⁶*10⁷*1.6*10⁻¹⁹ or >2,240 Coulombs/cm² while the data sheet for this tube specifies a 21% loss in quantum efficiency at 0.1 Coulomb/cm².

*3.2. Geometry:*

In order to measure the time of all protons in the silicon strip trackers, the TOF detector should be fully efficient starting at 0.5mm or less from the beampipe.

The TOF should also be compatible with both techniques used in forward physics to get close to the beam.

-In the Roman Pot technique a cylinder, whose bottom surface has a cutout for the beam, contains the detectors and is coupled to one side of the beampipe through a bellows. The bellows are compressed and the detectors enter the beampipe to within ~1mm from the beam.

-The alternative is to use the "Hamburg pipe". In this case a section of beampipe is machined away and replaced by a flat surface whose thickness is reduced close to the beam height. Detectors are placed against the flat surface and the whole assembly is pushed towards the beam.

*3.3. Segmentation:*

The high rates (0.5 tracks/bunch) require segmentation of the TOF since an event of interest is lost if either proton hits the same segment as another particle. For 7 segments the one arm efficiency is:

$$\mathit{eff}(1) = \frac{\sum_{n=1}^{\infty} P_m(n)\left(\frac{7-1}{7}\right)^{n-1}}{1 - P_m(0)}$$

where $P_m(n)$ is the Poisson distribution and the mean, m is =0.5.
Then the efficiency for exclusive events is $\mathit{Eff}^2 = 92\%$. At 220 m, due to the higher rates, it is 55%.

## 4. APD Detectors:

We have chosen devices with large signal (effective depth), gain and short risetime in response to charged particles. Advanced Photonics (AP) and Radiation Monitoring Devices (RMD) make similar structures which satisfy these requirements. We discuss our measurements on the RMD devices since relevant data exist on radiation damage.

Our design is based on an array of 7 identical APDs. Each has overall dimensions of 4mm*12.5 mm and an active area of 3mm*10mm so they are sensitive 0.5mm from the edge.

Our results are based on tests of a smaller, 20mm² AP detector and a 64mm² RMD detector.

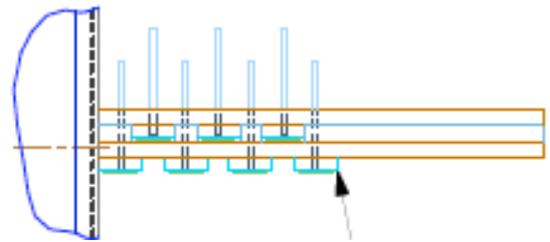

Figure 1. Top view of the 7 detectors starting at the left from the beam pipe. The detectors are staggered and overlap to eliminate the dead region.

## 5. Specifications:

Specifications for the APDs can be found in the manufacturer's data sheets [8]:

| Maximum Gain | ~10,000 |
|---|---|
| Capacitance | 0.7pF/mm² |
| HV (at G~1000) | 1650-1750 |
| Risetime | <1 nsec |

## 6. Measurements:

| N(electron-hole pairs) | 6000 |
|---|---|
| $\tau_R$(risetime) | 0.65 nsec |

| Pulse width | 5 nsec |
| --- | --- |
| f (excess noise factor) | 2.2 |
| $I_{surface}$ | 1,700 pA/mm$^2$ |
| $I_{Bulk}$ | 7 pA/mm$^2$ |
| $I_{bulk}(-30^oC)/I_{bulk}(22^oC)$ | 1/200 |

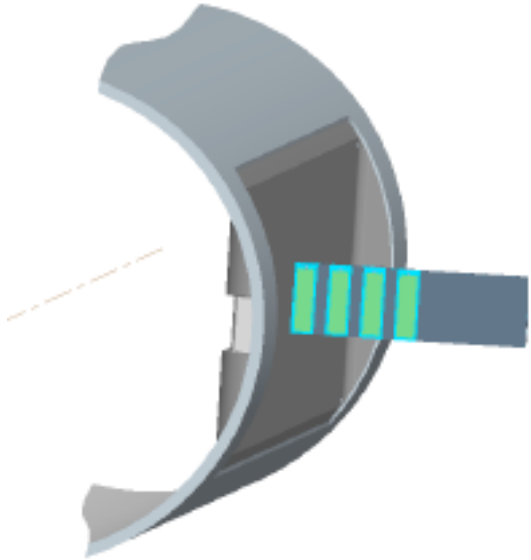

Fig.2. Detector configuration in the Hamburg pipe. The proportions of the pipe are distorted for purposes of illustration.

The number of electron-hole pairs was measured with a $^{106}$Ru β source using a gain of 1000.

Since pulse width is proportional to the APD capacitance (0.7 pF/mm$^2$), we interpolated between the larger and smaller devices and find 5 nsec. From this we calculate a maximum count rate of >100 MHz. This is more than adequate since the bunch collision rate is 40 MHz.

The limit on the integrated count rate is a radiation damage issue and is discussed below.

**7. Radiation**

*7.1. Annual Dose*

With a proton flux of 10 MHz we have an annual dose of $5*10^{13}$ (7 TeV protons)/cm$^2$ which is equivalent to $8.5*10^{12}$ neutrons/cm$^2$ or roughly 3% of the ATLAS silicon strip detector requirement.

While the background is probably uniform the signal peaks in the midplane so there is a higher dose in the midplane. Since the leakage current affects the whole detector, we consider the average dose.

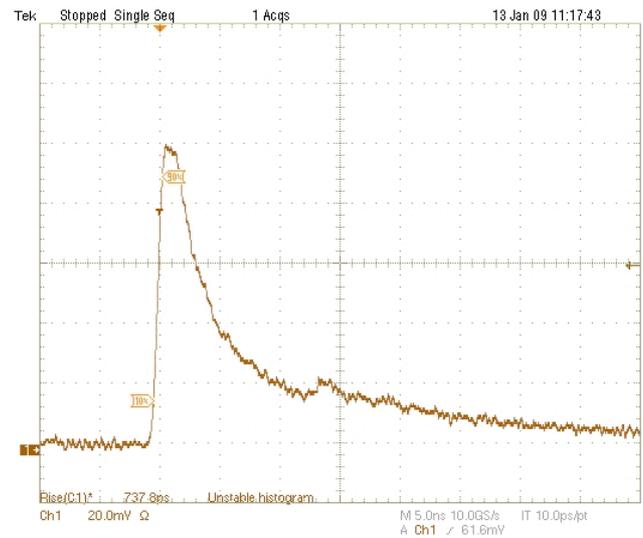

Figure 3. Signal from a 64mm$^2$ RMD APD. The scope is triggered by an $^{90}$Sr β passing through the APD and hitting a scintillator. From the 1 GHz bandwith of the oscilloscope we calculate an APD risetime of ~650 psec.

*7.2. Radiation Damage*

Radiation hardness of silicon devices is an issue for all aspects of the LHC- from the quench protection diodes in the accelerator to the inner layer of the ATLAS silicon strip tracker. In these examples the radiation hardness requirement is $3*10^{13}$ and $3*10^{14}$ neutrons/cm$^2$.

The main effect on silicon devices is displacement damage [9], which increases the bulk leakage current in proportion to the dose.

Effective dose is proportional to the non-ionizing energy loss (NIEL). For a given radiation type the NIEL [10] is used to calculate the 1 MeV neutron equivalent dose.

The dose limit for our APD hasn't yet been measured. Performance as a photodetector was measured by RMD up to $5*10^{12}$ neutrons/cm$^2$ at which point there was an 80% loss in quantum efficiency [11]. This is irrelevant for use as a charged particle detector.

Below we argue that we are unlikely to see any significant degradation in performance as a time of flight counter.

*7.3. Radiation tests of APDs:*

RMD observed no change in gain or signal but an increase in leakage current at $5*10^{12}$. When the detector was cooled to $-30^oC$ the noise was identical to that of the unirradiated detector.

CMS [11] confirmed the linearity of leakage current vs. dose in APDs up to $2*10^{14}$ n/cm$^2$. So after a dose of $8.5*10^{12}$ the leakage current of the RMD detector is increased by less than a factor of $(8.5/5)*200$ since the drop in current due to temperature is a factor of 200.

CMS [12] also measured damage coefficients for several types of APDs and found $\alpha \sim 1.2*10^{-16}$ A/(n-cm), where

$$I^{APD}_{dark}=\alpha*L_{eff}*Area*Dose$$

so with $L_{eff}\sim 60\mu m$ (i.e. 6000e/(100e/$\mu$m)) we find
$I=1.2*10^{-16}*(60*10^{-4})*0.3*8.5*10^{12}=1.9\mu A$
or 10 nA at $-30^oC$

*7.4. Time Resolution:*

With leading edge timing the noise jitter is [1]
$\delta t=\sigma(e)/(d(Amplitude)/dt)$
where $d(Amplitude)/dt = N_{e-h}/\tau_R$
and $\tau_R$ is the signal risetime. In our case the dominant noise contribution is from leakage current:
$\sigma_{leakage}^2(e)=(I_{surface}/G^2+I_{bulk}*f)* \tau_{shaping}/q_e$

so the noise limit to the time jitter at $-30^oC$ and after 10 years of irradiation is
$\delta t=48e/6000e* \tau_R= 5.2$ psec

As mentioned before, the amplitude jitter, which has contributions from gain nonuniformity(<3%), excess noise factor (<2%) and the Landau distributed fluctuation of energy loss (18%) totals 18%. We therefore need to correct for walk to a level of 0.18*650 psec/10 psec or ~11. This is roughly what was achieved in ref. [3].

**8. The 220 m Location:**

The count rate is 6 times larger at 220 m and the dose rate is 3 times larger than the 420 m location considered above. A smaller dead zone near the beampipe would significantly improve the physics performance so we are investigating options for reducing the dead zone.

**9. Conclusions:**

This APD based design addresses all of the challenges of high rate, high resolution time of flight measurement in forward physics at the LHC.

We will do further studies of timing performance in time of flight measurements. We plan also to expose our APDs to a dose of several LHC years equivalent.

**10. Acknowledgements:**

We would like to thank Dick Farrell, Iouri Musienko, Steve Reucroft, Valentina Avati, Hubert Niewiadomski and Nikolai Mokhov for helpful discussions.